\documentclass[12pt]{emulateapj}
\usepackage{ulem}
\usepackage{natbib}
\usepackage{color}
\usepackage{amssymb}
\usepackage{times}
\usepackage{pifont}
\usepackage{rotating}
\usepackage{amsmath}
\usepackage{graphicx}
\usepackage{pslatex}
\usepackage{epstopdf}
\usepackage{hyperref}

\def\be{\begin{eqnarray}}   \def\ee{\end{eqnarray}}
\def\ben{\begin{equation}\begin{aligned}} \def\een{\end{aligned}\end{equation}}

\def\sec#1{Section~\ref{sec:#1}}
\def\fig#1{Figure~\ref{fig:#1}}
\def\figs#1{Figures~\ref{fig:#1}}
\def\tab#1{Table~\ref{tab:#1}}

\definecolor{grey}{rgb}{0.35,0.35,0.35}

\begin{document}
\title{Stellar population synthesis
based modelling of the Milky Way using asteroseismology of dwarfs and subgiants from Kepler.}
\author{Sanjib Sharma\altaffilmark{1}, Dennis Stello\altaffilmark{1,2}, Daniel Huber \altaffilmark{1,2,3}, Joss  Bland-Hawthorn\altaffilmark{1}, Timothy R.  Bedding\altaffilmark{1,2}}
\altaffiltext{1}{Sydney Institute for Astronomy, School of Physics,
University of Sydney, NSW 2006, Australia}
\altaffiltext{2}{Stellar Astrophysics Centre, Department of
Physics and Astronomy, Aarhus University, DK-8000 Aarhus
C, Denmark}
\altaffiltext{3}{SETI Institute, 189 Bernardo Avenue, Mountain View, CA 94043, USA}
\begin{abstract}
Early attempts to apply asteroseismology to study the Galaxy have already
shown unexpected discrepancies for the mass distribution of stars between
the Galactic models and the data; a result that is still unexplained.
Here, we revisit the analysis of the asteroseismic sample of dwarf
and subgiant stars observed by {\it Kepler} and investigate
in detail the possible causes for the reported discrepancy.
We investigate
two models of the Milky Way based on stellar population synthesis,
{\sl Galaxia} and TRILEGAL.
In agreement with previous results, we find that TRILEGAL
predicts more massive stars compared to {\sl Galaxia}, and that
TRILEGAL predicts too many blue stars compared to 2MASS
observations.
Both models fail to match the distribution of the
stellar sample in $(\log g,T_{\rm eff})$ space, pointing to
inaccuracies in
the models and/or the assumed selection function.
When corrected for this mismatch in $(\log g,T_{\rm eff})$ space,
the mass distribution calculated by {\sl Galaxia} is broader
and the mean is shifted toward lower masses compared to that of
the observed stars. This behaviour is similar to what has
been reported for the {\it Kepler} red giant sample.
The shift between the mass distributions is equivalent
to a change of 2\% in $\nu_{\rm max}$, which
is within the current uncertainty in the
$\nu_{\rm max}$ scaling relation.
Applying corrections to the $\Delta \nu$
scaling relation predicted by the stellar models
makes the observed mass distribution significantly
narrower, but there is no change to the mean.
\end{abstract}

\section{Introduction}
Our current understanding of the Milky Way has, to a large extent, been informed
by stellar data from large scale photometric, astrometric, and
spectroscopic surveys, such as,
2MASS \citep{2006AJ....131.1163S},
SDSS \citep{2008ApJ...673..864J},
Hipparcos \citep{1997yCat.1239....0E},
GCS \citep{2004A&A...418..989N},
RAVE \citep{2013AJ....146..134K},
SEGUE \citep{2009AJ....137.4377Y},
APOGEE \citep{2013AJ....146...81Z}, and
Gaia-ESO \citep{2012Msngr.147...25G}.
As a result, we have already come a long way from simple empirical models
of the Galaxy that fit star counts in a few lines of sight
\citep{1980ApJS...44...73B,1980ApJ...238L..17B,1984ApJS...55...67B}, to
models that aim to be dynamically self-consistent
\citep{2003A&A...409..523R,2010MNRAS.401.2318B,2012MNRAS.426.1328B,2011MNRAS.413.1889B,2012MNRAS.426.1324B,2012MNRAS.419.2251M,2014A&A...564A.102C,2009MNRAS.396..203S,2009MNRAS.399.1145S,2014ApJ...793...51S}.
Some of these models, such as , {\sl Besan\c{c}on} \citep{2003A&A...409..523R},
TRILEGAL \citep{2005A&A...436..895G}, and {\sl Galaxia}
\citep{2011ApJ...730....3S} have been constructed to directly satisfy
the observational constraints from various large scale photometric,
astrometric and spectroscopic surveys of the Milky Way.
However, to understand the Milky Way's formation history,
and hence to further verify the models,  it
is important to know the fundamental properties of the stars, including
radius and mass.  Until recently, it has been difficult to reliably determine
these properties model-independently for large numbers of distant stars.
Fortunately, the space missions
{\rm CoRoT} \citep{2006ESASP1306...11B} and {\it Kepler}
\citep{2010Sci...327..977B},
and now also K2 \citep{2014PASP..126..398H},
provide highly accurate time-series photometry of thousands of stars across
the Galaxy, from which we can obtain asteroseismic information that is
sensitive to, and hence capable of measuring, stellar radius and mass in a
largely model independent way.

A promising approach to take advantage of the large ensembles of
seismically-inferred stellar properties is to invoke stellar population
synthesis-based models of the Milky Way
\citep[e.g.~][]{2009A&A...503L..21M,2011Sci...332..213C,2016ApJ...822...15S}.
This offers
a way to link stellar structure and evolution with Galactic structure and
evolution by combining isochrones with star-formation
history, the initial mass
function, and the spatial distribution of stars of the
Galaxy. This allows one to predict
stellar observables like temperature, photometry, asteroseismic parameters,
as well as fundamental stellar properties such as radius and mass.

\citet{2011Sci...332..213C} used
the first seven months of {\it Kepler} data of dwarfs and
subgiants, to compare
the distributions of seismically-inferred radii and masses of about 400 stars
with a synthesis-based Galactic model using TRILEGAL. They found that the radius
distribution of the synthetic population matched the data, but
the mass distribution significantly under-predicted
the number of low-mass stars
($M<1.15 {\rm M}_{\odot}$), and hence over-predicted the number of more
massive (younger) stars. Using red
giants from {\it Kepler}, we found the opposite effect when
comparing the observed masses with predictions from
the Galactic model {\sl Galaxia} \citep{2016ApJ...822...15S}.
We showed that in the {\it Kepler} region for a magnitude
limited sample, TRILEGAL
predicts more blue stars as compared to 2MASS, while
{\sl Galaxia} has no such problem. Because blue stars are
young and massive, this suggests that TRILEGAL
overpredicts the number of young and massive stars.
Hence, it is important to compare
the observed masses of dwarfs and subgiants
with the predictions from {\sl Galaxia}.

Besides inaccuracies in the Galactic model, there are a number
of other factors that could contribute to the
mismatch in  the mass distributions seen by \citet{2011Sci...332..213C}.
(i) Inaccuracies in the selection function can lead to a
mismatch and need to be checked.
There could be systematics associated with the
algorithm used to estimate average seismic parameters.
(ii) The probability to detect oscillations can differ from one
algorithm to another and this can lead to differences
in the selection function.
Since the analysis by \citet{2011Sci...332..213C}, a new
data set  of dwarfs and subgiants has been published
by \citet{2014ApJS..210....1C}, which
used a  different algorithm to estimate the
seismic parameters.
Additionally, it contains more stars and extends to
slightly lower gravities.
Hence, it is necessary to investigate the mass distributions with the
new data set as well.
(iii) Theoretical modelling of the stellar oscillations
predict departures from the $\Delta \nu$ scaling
relation \citep{2009MNRAS.400L..80S,2011ApJ...743..161W,2013EPJWC..4303004M}
and the effect of this needs to be
taken into account.
(iv) To estimate mass from average seismic parameters,
one has to adopt certain solar reference values. Currently,
there is no consensus on the choice of these with systematic
differences ranging from 1\% to 2\%. Hence, one needs to investigate
whether the discrepancy between observations and predictions
is less or greater than the current diversity in the
solar reference values.

In this paper we revisit the dwarf/subgiant
analysis of \citet{2011Sci...332..213C}
and analyze each of the above mentioned factors.
In Section 2, we discuss the observational data and the
Galactic models. Systematics associated with the different
data sets and Galactic models are discussed here.
In Section 3, we analyze the asteroseismic information for
different data sets and different Galactic models and
discuss  the role of the selection function.
We also do a quantitative study of the difference
between observed and predicted mass distributions.
Finally, in Section 4 we discuss and conclude our findings.

\section{Data, scaling relations and Galactic models}
\begin{figure}
\centering \includegraphics[width=0.46\textwidth]{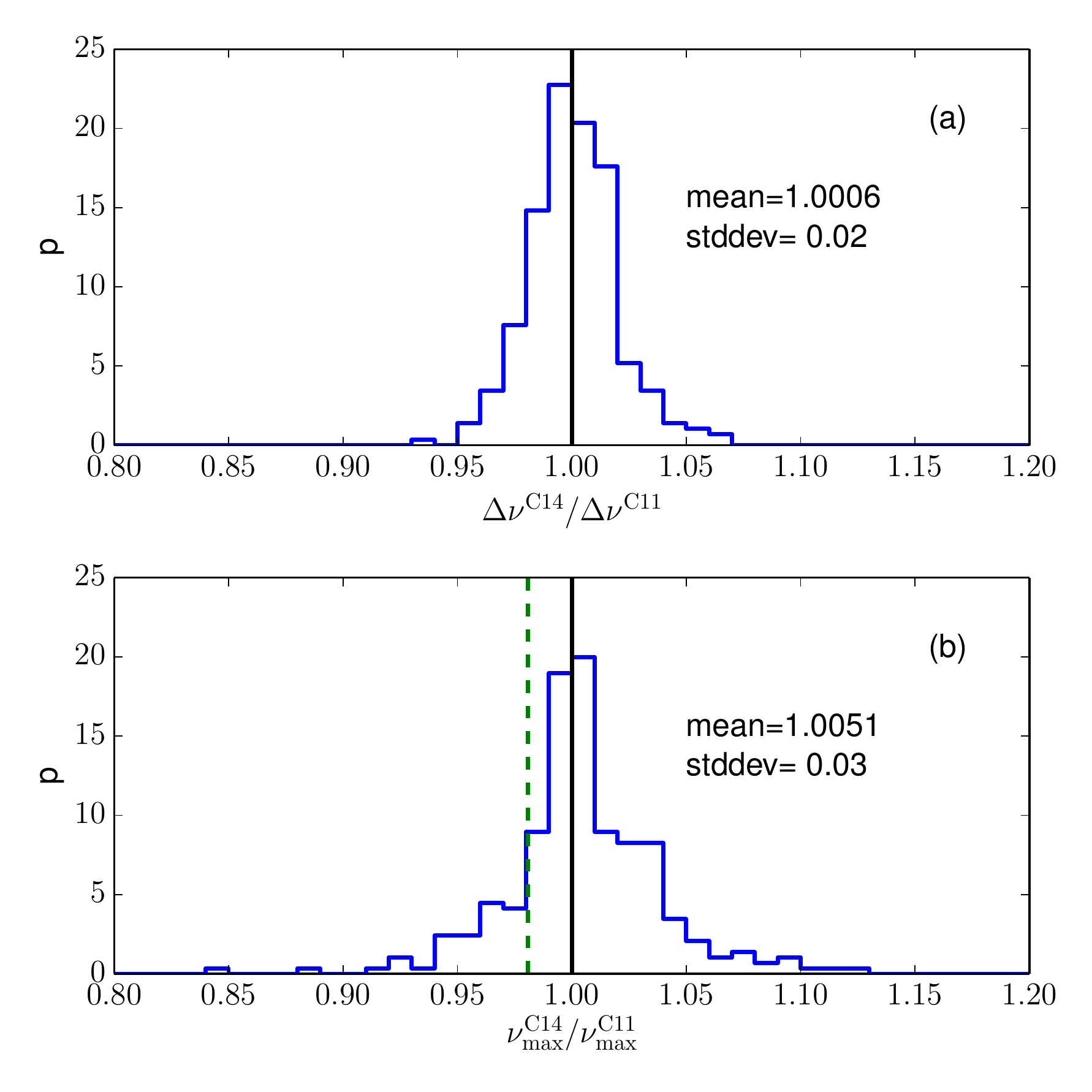}\caption{(a) Ratio of $\Delta
\nu$ for the 290 stars in common
between the  Chaplin-2014 and
Chaplin-2011 samples. (b) Same as panel (a) but for
$\nu_{\rm max}$.   The dashed line
in panel (b) shows  the ratio of $\nu_{\rm max,\odot}$
adopted in Chaplin-2014 to that adopted in  Chaplin-2011.
There is no systematic shift between
the two  methods. The standard deviation is of the order 
of the uncertainties on the estimated values of 
$\nu_{\rm max}$ (4\%) and $\Delta \nu$ (2\%). 
\label{fig:c14_c11_nu}}
\end{figure}
\subsection{Observational data}
The observed asteroseismic information is in the form of
the average seismic parameters $\Delta \nu$ (average frequency spacing
between overtone oscillation modes) and $\nu_{\rm max}$
(frequency at maximum oscillation power)
that are extracted from time series photometry using a
specific algorithm/method.
Prior to the launch of {\it Kepler}, about 2000
stars with $Kp < 12$ mag were selected as potential
asteroseismic dwarf and subgiant targets based on their
parameters in the Kepler Input Catalog (KIC)
\citep{2011AJ....142..112B}.  They were all observed with
short cadence for one
month each during an initial 10-month seismic survey phase.
Four hundred stars showed detectable oscillations after the first
seven months and were presented by \citet{2011Sci...332..213C}
(hereafter denoted as Chaplin-11 sample).
Following the completion of the 10-month survey, an
updated dwarf/subgiant sample (518 stars) showing
oscillations was presented by \citet{2014ApJS..210....1C}
(here after denoted as Chaplin-14 sample), where a
method different from that of \citet{2011Sci...332..213C}
was used for extracting seismic parameters.
In the Chaplin-14 sample 467 stars have
measured values of $\Delta \nu$, $\nu_{\rm max}$ and $T_{\rm  eff}$.
Of these, 290 stars are in common with the
previous Chaplin-11 sample.  In \fig{c14_c11_nu}, we
compare the $\Delta \nu$ and $\nu_{\rm max}$ values of
Chaplin-2014 with those of Chaplin-2011, for
stars common to both samples.  It can be seen that there are 
no systematic shift between the two data sets.  This means
there is no method-specific differences in the two data
sets. In this paper we mainly make use of the
Chaplin-2014 sample, because it has more stars.

\subsection{Scaling relations and solar reference values}\label{sec:astero}
The stellar mass and
radius can be estimated from the seismic parameters $\Delta \nu$ and
$\nu_{\rm max}$, and the effective temperature $T_{\rm eff}$
using the following scaling relations:
\be
\label{equ:scaling_m}
\frac{M}{\rm {\rm M}_\odot} &  = &  \left(\frac{\nu_{\rm
max}}{f_{\nu_{\rm max}} \nu_{\rm max,
\odot}}\right)^{3}\left(\frac{\Delta \nu}{f_{\Delta \nu}
\Delta \nu_{\odot}}\right)^{-4}\left(\frac{T_{\rm eff}}{T_{\rm
eff, \odot}}\right)^{1.5}   \\
\label{equ:scaling_r}
\frac{R}{\rm R_\odot}  & = & \left(\frac{\nu_{\rm
max}}{f_{\nu_{\rm max}} \nu_{\rm max, \odot}}\right)\left(\frac{\Delta
\nu}{f_{\Delta \nu} \Delta \nu_{\odot}}\right)^{-2}\left(\frac{T_{\rm
eff}}{T_{\rm eff, \odot}}\right)^{0.5}.
\ee
These relations are based on the relations $\Delta \nu \propto \rho^{1/2}$, and
$\nu_{\rm max} \propto g/T_{\rm eff}^{1/2}$
\citep{1991ApJ...368..599B,1995A&A...293...87K,2011A&A...530A.142B} 
, which in turn are based on the assumption that the structure of any 
given star is homologous with respect to the Sun. This assumption
is not strictly correct and can lead to departures
from the scaling relations. To accommodate these   
departures 
we have introduced the factors $f_{\nu_{\rm max}}$ and
$f_{\Delta \nu}$. 
There is also considerable uncertainty regarding 
the choice of solar reference values and this 
leads to uncertainties in $f_{\nu_{\rm max}}$ and $f_{\Delta
  \nu}$, when we adopt a specific set of canonical solar 
reference values. Below, we discuss this in detail.

\begin{table}
\caption{Solar reference values for different methods
of computing average seismic parameters.}
\begin{tabular}{l l l} \hline
Method & $\Delta \nu_{\odot} (\mu {\rm Hz})$ & $\nu_{\rm
max,\odot} (\mu {\rm Hz})$ \\ \hline
SYD \tablenotemark{a} &$135.10 \pm 0.01  $  & $3090 \pm 3$ \tablenotemark{e}
\\ \hline
CAN \tablenotemark{b} &   $134.88 \pm 0.04 $  & $3120 \pm 5  $ \\ \hline
COR \tablenotemark{c} &  $134.90 \pm 0.1  $  & $3060 \pm 10 $ \\ \hline
OCT \tablenotemark{d} &  $135.03 \pm 0.1  $  & $3140 \pm 10 $ \\ \hline
\end{tabular}
\tablenotetext{1}{\citet{2009CoAst.160...74H,2011ApJ...743..143H,2013ApJ...767..127H}}
\tablenotetext{2}{\citet{2010A&A...522A...1K}}
\tablenotetext{3}{\citet{2012A&A...537A..30M,2013A&A...556A..59H}}
\tablenotetext{4}{\citet{2013A&A...556A..59H}}
\tablenotetext{5}{\citet{2011ApJ...743..143H} report an uncertainty
of $30\ \mu {\rm Hz}$ for $\nu_{\rm
max,\odot}$ and $10\ \mu {\rm Hz}$ for $\Delta \nu_{\odot}$. However, this is for one 30 day solar time
series subset out of
111 analyzed by them in total.}
\label{tab:tb1}
\end{table}

It is clear from the scaling relations that to estimate mass
and radius one has to adopt some
solar reference values, $\Delta \nu_{\odot}$ and $\nu_{\rm
max,\odot}$. Unfortunately, there is no consensus on the
choice of solar reference values. Ideally, to estimate them
we would require high quality data of the Sun in the {\it Kepler}
bandpass, which, unfortunately is not available. The data for the
Sun is available in other bandpasses and this has been
analyzed. The results using  the
SOHO/VIRGO green channel data \citep{1997SoPh..170....1F}
are shown in \tab{tb1} for various methods.
While the estimates of $\Delta \nu_{\odot}$ agree (difference
about 0.2\%), the
estimates of $\nu_{\rm max,\odot}$ differ significantly
(difference about 2.5\%) and so far this disagreement has not
been explicitly explained.
The most likely cause for the differences is the
method-specific systematics associated  with the estimation
of $\nu_{\rm max}$.
This would argue for the
use of ``method-specific'' solar values, meaning the
values returned from solar data when using the same method
(pipeline) as used for the rest of the stellar sample.
However, there is no strong evidence to back up the use
of method-specific values.
On the contrary, \citet{2013A&A...556A..59H} found
that for red giants, method-specific solar reference values
introduce biases.  In other words, method-specific systematics
in $\nu_{\rm max}$ for the Sun are not necessarily
representative of the systematics for other stars.

Similar to giants, for dwarfs and subgiants 
it is not clear whether one should adopt method
specific  solar reference values.
The Chaplin-14 sample adopted the SYD method
\citep{2009CoAst.160...74H,2011ApJ...743..143H}
for computing the seismic parameters.
The solar reference values
corresponding to this method are
$\Delta \nu_{\odot}=135.1\ \mu{\rm
Hz},\  \nu_{\rm max,\odot}=3090\ \mu {\rm Hz}$.
For the Chaplin-11 sample, however, a
different method was used, namely the OCT pipeline
\citep{2013A&A...556A..59H} available at that time 
whose method specific solar reference 
values were $\Delta \nu_{\odot}=134.9\ \mu {\rm Hz}$ and
$\nu_{\rm max,\odot}=3150\ \mu {\rm Hz}$. 
These were adopted by \citet{2011Sci...332..213C} 
for computing the stellar masses in their sample.
As discussed earlier (\fig{c14_c11_nu}), there is no
systematic shift in seismic parameters between the Chaplin-11 and 
Chaplin-14 data sets.  
Hence, $\nu_{\rm max,\odot}=3090\ \mu{\rm Hz}$ is an
equally valid choice for estimating masses.
For clarity throughout, we adopt
$\Delta \nu_{\odot}=135.1\ \mu{\rm  Hz}$,
and   $\nu_{\rm max,\odot}=3090\ \mu {\rm Hz}$.

Based on the discussion in the previous two paragraphs,  we
conclude that $f_{\nu_{\rm max}}$ is uncertain
by at least 2\%, and we will use this fact later.

\begin{figure}
\centering \includegraphics[width=0.46\textwidth]{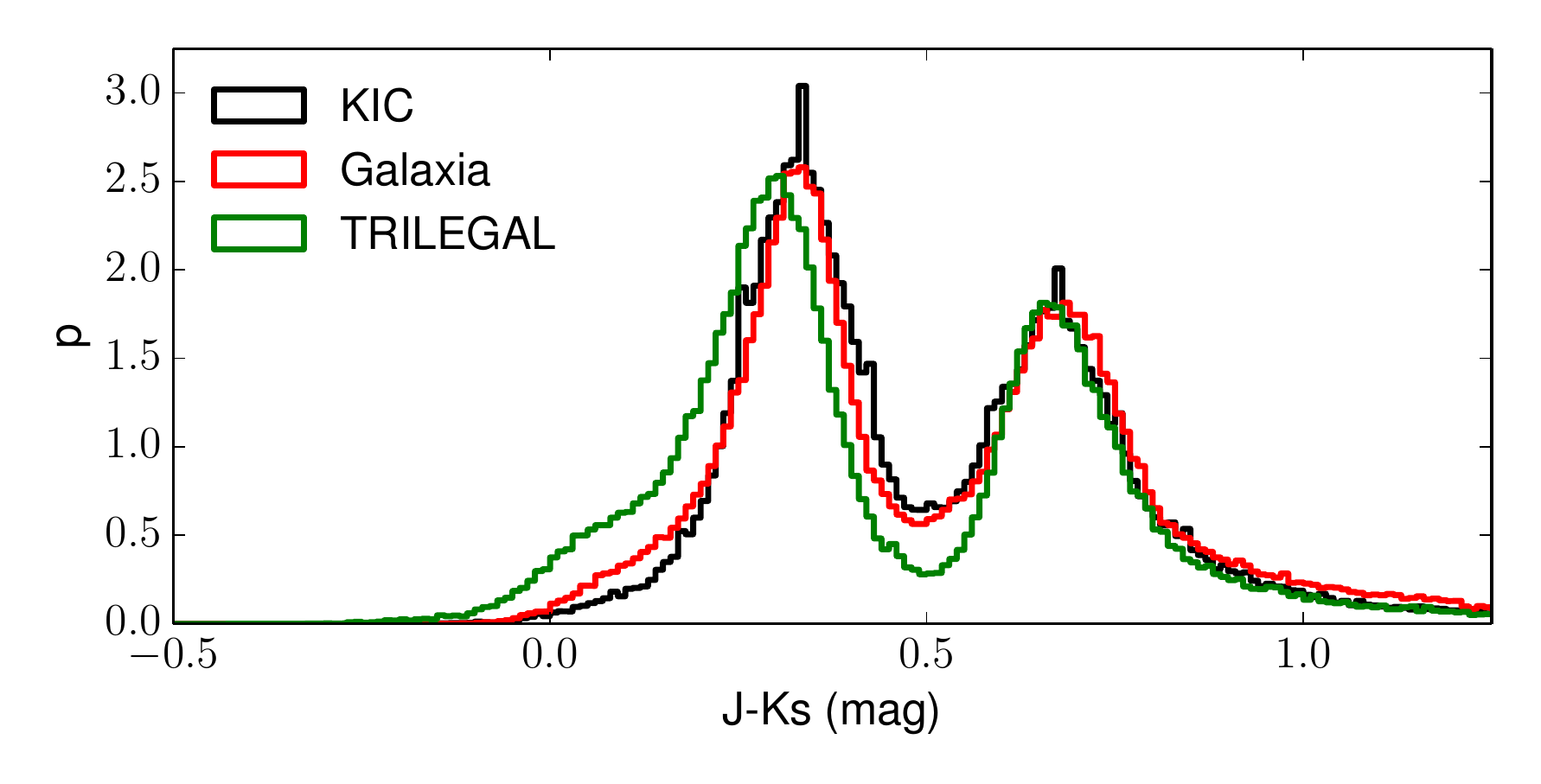}\caption{(a) $J-K_s$ color distribution of stars with $r<14$
in the KIC (black)  compared with predictions from {\sl
Galaxia} (red) and TRILEGAL (green).
(b) as panel (a), but for $g-r$ color. The
integrated probability distributions are scaled to unity.
\label{fig:galaxia_trilegal_jkgr}}
\end{figure}
\begin{figure*}
\centering \includegraphics[width=0.96\textwidth]{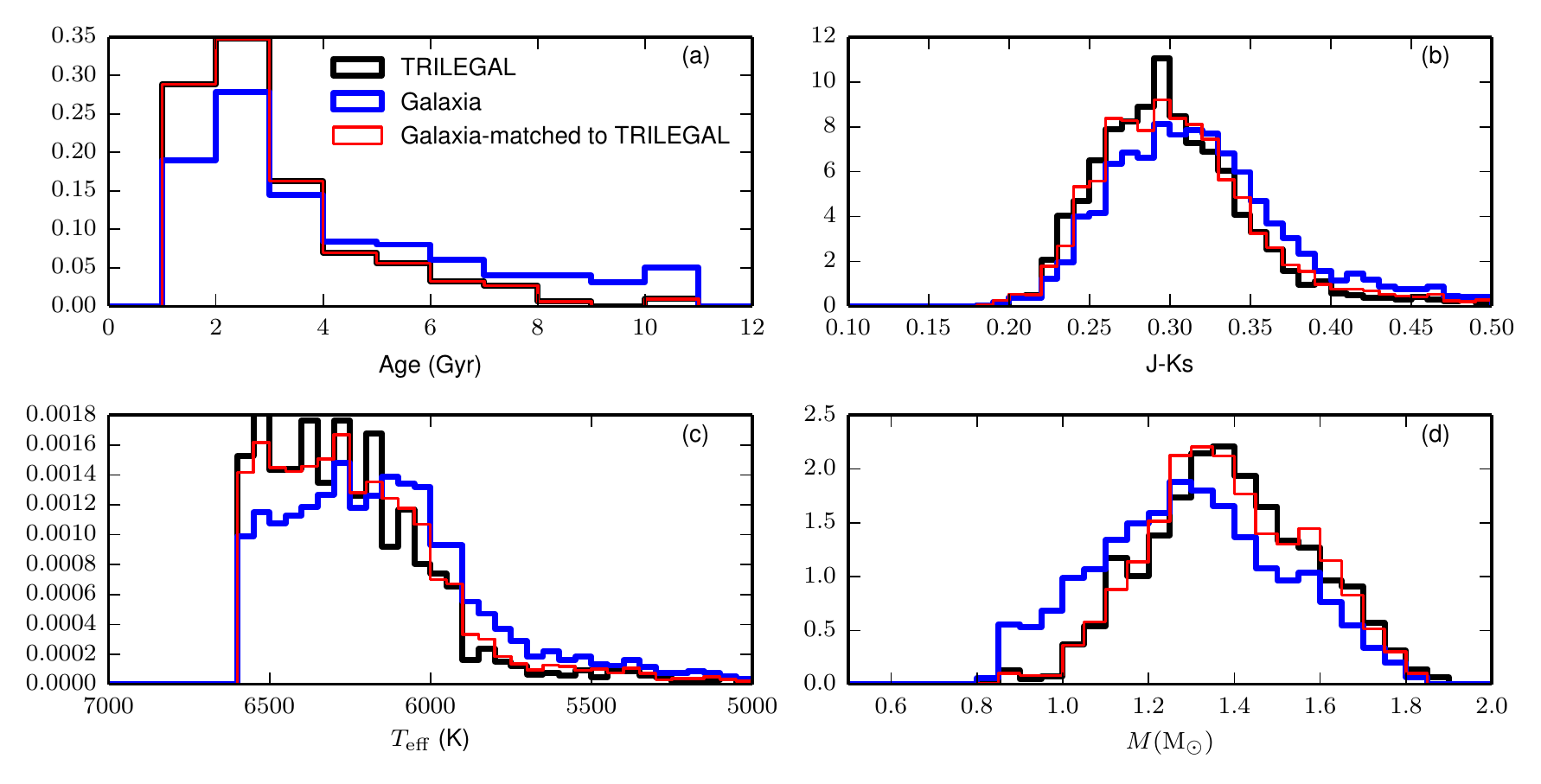}\caption{Comparison of stellar properties of TRILEGAL
(black) and {\sl Galaxia} (blue). Shown are
distributions of age (panel a), color (panel b),
temperature (panel c), and mass (panel d).  The curves in red
show the predictions of {\sl Galaxia} when its stars are
reweighted to match the age distribution of the TRILEGAL
stars. In panel a, the red and black lines are on top of
each other.
\label{fig:subgiants_summary}}
\end{figure*}

\subsection{Galactic models} \label{sec:models}
The main Galactic stellar-population-synthesis model
used in this paper is from the {\sl Galaxia}\footnote{\url{http://galaxia.sourceforge.net}} code
\citep{2011ApJ...730....3S}. It uses a Galactic model
that is based on the {\sl Besan\c{c}on} model
by \citet{2003A&A...409..523R} but with some modifications.
{\sl Galaxia} has a 3D extinction scheme that is based on
\citet{1998ApJ...500..525S} dust maps.
We also  apply a low latitude correction to the
dust maps as in \citet{2014ApJ...793...51S}.
The isochrones used to predict the stellar
properties are from the Padova database
\citep{1994AAS..106..275B,2008A&A...482..883M}. The unique feature of {\sl
Galaxia} is its novel star-spawning scheme which,
unlike previous codes, does not discretize the spatial
dimensions into multiple lines of sight; instead  it
generates a continuous three-dimensional distribution of
stars.

For comparison, we also used the
TRILEGAL\footnote{\url{http://stev.oapd.inaf.it/cgi-bin/trilegal}}
Galactic stellar-population-synthesis model
\citep{2005A&A...436..895G}. TRILEAGL as a code is very
flexible and offers multiple options to the user to change
various aspects of the Galactic model, i.e.,  
IMF, SFR, age scale height, local surface density of stars 
and so on. 
However, there is a default version of the 
Galactic model that is advocated in the  
\citep{2005A&A...436..895G} paper and is commonly used. 
We here wish to investigate the
discrepancies reported by \citet{2011Sci...332..213C}, who 
do not mention any specific changes to the default
settings. So we choose to use the default model of
TRILEGAL for our analysis. 

Unlike {\sl Galaxia}, TRILEGAL
cannot generate stars over a wide angular area, so we
generated stars along 21 lines of sight pointing towards the
centers of the 21 {\it Kepler} CCD-modules.
We used the default settings but with binary stars turned off.
To model the dust, we used the 3D extinction model of {\sl Galaxia}.
One noticeable difference between {\sl Galaxia} and
TRILEGAL is that {\sl Galaxia} uses a constant star-formation
rate, while the default setting in TRILEGAL
uses a two-step star-formation rate, which
is twice as large between 1-4 Gyr as at any other time.

\subsubsection{Comparing Galactic models with the Kepler
Input Catalog} \label{kic_compare}
Before comparing the Galactic models with the data, it is
important to check
that they reproduce the stellar photometry in the
KIC.  This is because the KIC formed the basis for the selection of the
observed sample of stars.  Hence, any significant mismatch between the
models and the KIC would indicate a fundamental problem with
the models,
making it difficult to perform meaningful comparisons with
the seismic data.

In \citet{2016ApJ...822...15S}, we analyzed the
$(J-K_s)$ color distribution of a magnitude limited sample
($r<14$ mag) of stars in the {\it Kepler} field and below
we summarize the results for the completeness of this paper.
We generated two synthetic catalogs of the Milky Way, using
each of {\sl Galaxia} and TRILEGAL. Stars from the synthetic
catalogs and from the
KIC that lay within 8 degrees of the center of the {\it Kepler} field
and with magnitude $r<14$ were selected for comparison.
In the KIC, the repeatability of photometry for stars brighter than
magnitude 14 is about 2\%, so it is very likely to be
complete for $r<14$.
We show the resulting distributions of $J-K_s$ color in
\fig{galaxia_trilegal_jkgr}.
The prediction of {\sl Galaxia} is in excellent agreement
with the stars from the KIC, but that from TRILEGAL shows
a significant difference.  TRILEGAL correctly reproduces the region around
the red peak of the color distribution, but not around the blue peak.
Specifically, it overestimates the number of stars to the
left of the blue peak and
underestimates the number of stars on the right side of the blue
peak. Stars leftward of the blue peak are typically young main-sequence
stars, suggesting that TRILEGAL over predicts the number of
young stars in the {\it Kepler} field.

Now, to better understand the above mentioned difference between TRILEGAL and
{\sl Galaxia}, we selected stars from both models
satisfying $4000 {\rm\ K} <T_{\rm eff}< 6600 {\rm\ K}$,
$3.7<\log g<4.2$ and $r< 12$ mag (we call this selection
criteria $S_{\rm dwarf}$). This was designed to
select dwarfs and subgiants, which are the main focus of
this paper. The distributions of stellar properties are shown
in \fig{subgiants_summary}.
The same color difference as seen in
\fig{galaxia_trilegal_jkgr} can be seen here. A difference
in age, temperature and mass distributions can also be
seen.
Mass, age, and metallicity are the three
intrinsic properties of a star that largely define
its observable properties.
The metallicity distributions (not shown here)
did not show any significant difference.
Hence,  the color
difference is most likely due to differences in mass and
age distributions. 
Next, we therefore investigate if the color,
  temperature, and mass distributions would match
(between {\sl Galaxia} and TRILEGAL) if we were to  
alter the Galactic model of {\sl Galaxia} in such a 
way that the age distribution of stars obeying the 
subgiant and dwarf selection criteria matches the 
TRILEGAL prediction. 
However we cannot easily alter the
model in {\sl Galaxia}, instead   
we use the idea of importance sampling. We assign a weight to 
each {\sl Galaxia} star such that the weighted age distribution 
matches that of TRILEGAL. Using these weights we then  
compute the weighted distriutions of other quantities like 
color, temperature, and mass, and compare them with those 
of TRILEGAL.    
We found that the color,
temperature and mass distributions of the reweighted 
{\sl Galaxia} sample now matched the TRILEGAL
sample.  Reweighting the {\sl Galaxia} sample to match
the mass distribution of the TRILEGAL sample produced
similar results. This is expected because
mass and age are correlated for the types of
stars that we analyze.
This shows that differences between {\sl Galaxia} and
TRILEGAL are mainly related to age and/or mass, and
not due to differences in isochrones. If the difference
were due to isochrones, then even after forcing a match on
mass, age and metallicity, the two models would have
shown differences in the color distribution.
The two main
factors that control the age distribution of a stellar 
sample are the star-formation rate and the age scale height
relation. Both of these factors are different between  
{\sl Galaxia} and TRILEGAL and could be 
responsible for the mismatch in the $J-K_s$ color 
distribution.

To conclude, we find that TRILEGAL cannot fit the color
distribution of the stars, probably because it predicts
too many young and massive stars. Unless we can explain
the mismatch in the color distribution by some other means
(e.g., systematics in the isochrones), a color or
temperature-limited sample of dwarfs and subgiants selected from
TRILEGAL is expected to be biased towards higher masses.

\begin{figure}
\centering \includegraphics[width=0.5\textwidth]{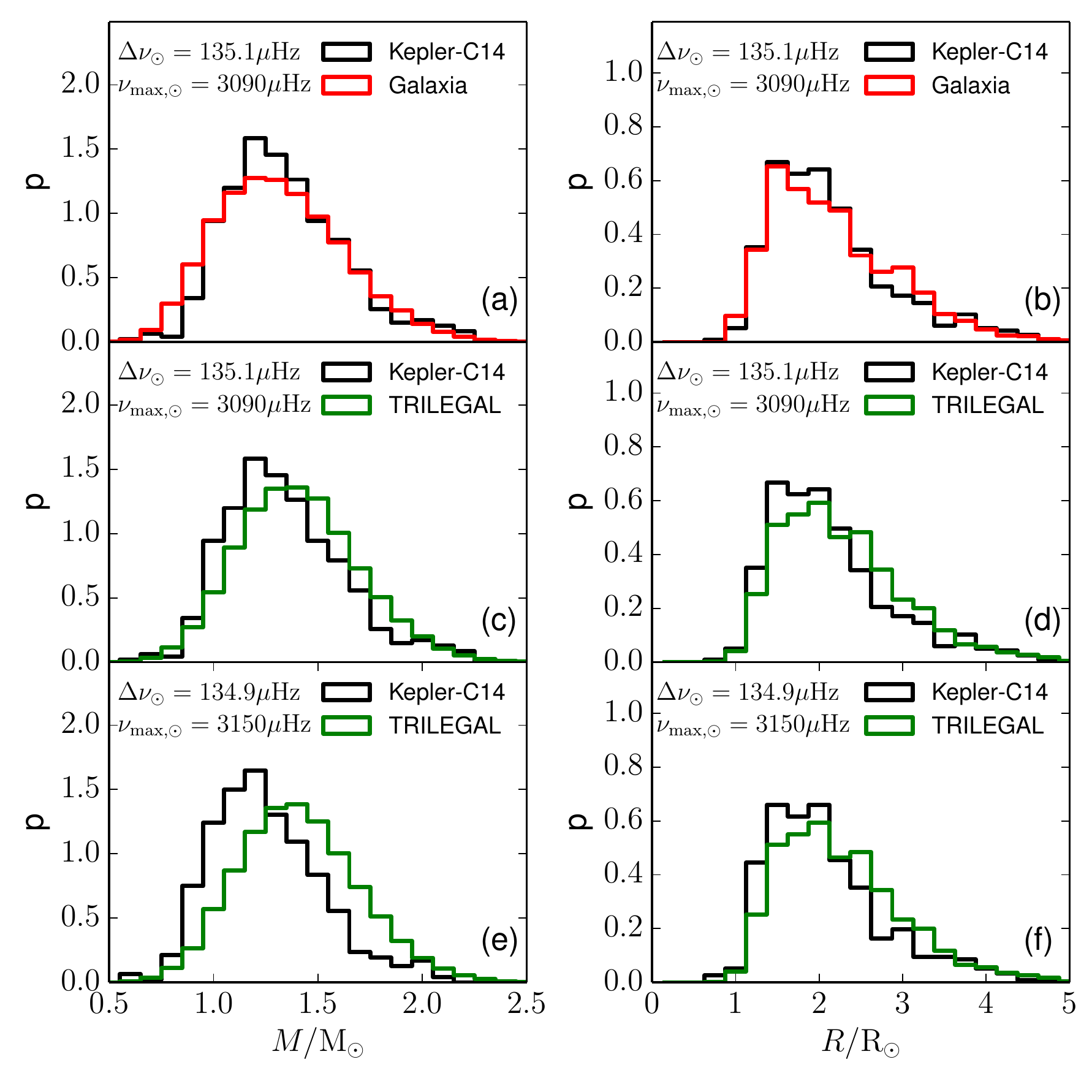}\caption{Mass and radius distributions for Chaplin-14 sample observed by
{\it Kepler} (black) and predicted by {\sl Galaxia}
(red) and TRILEGAL (green).
(a-b): Result using {\sl Galaxia} and our choice of solar
reference values for $\Delta \nu$ and $\nu_{\rm max}$.
(c-d): Result using TRILEGAL  and our choice of solar
reference values.
(e-f): Result using TRILEGAL and the solar reference values used by
\citet{2011Sci...332..213C} (our reproduction of the Chaplin-11 result).
The integrated probability distributions are scaled to unity.
\label{fig:c14s0_mr}}
\end{figure}

\begin{figure}
\centering
\includegraphics[width=0.5\textwidth]{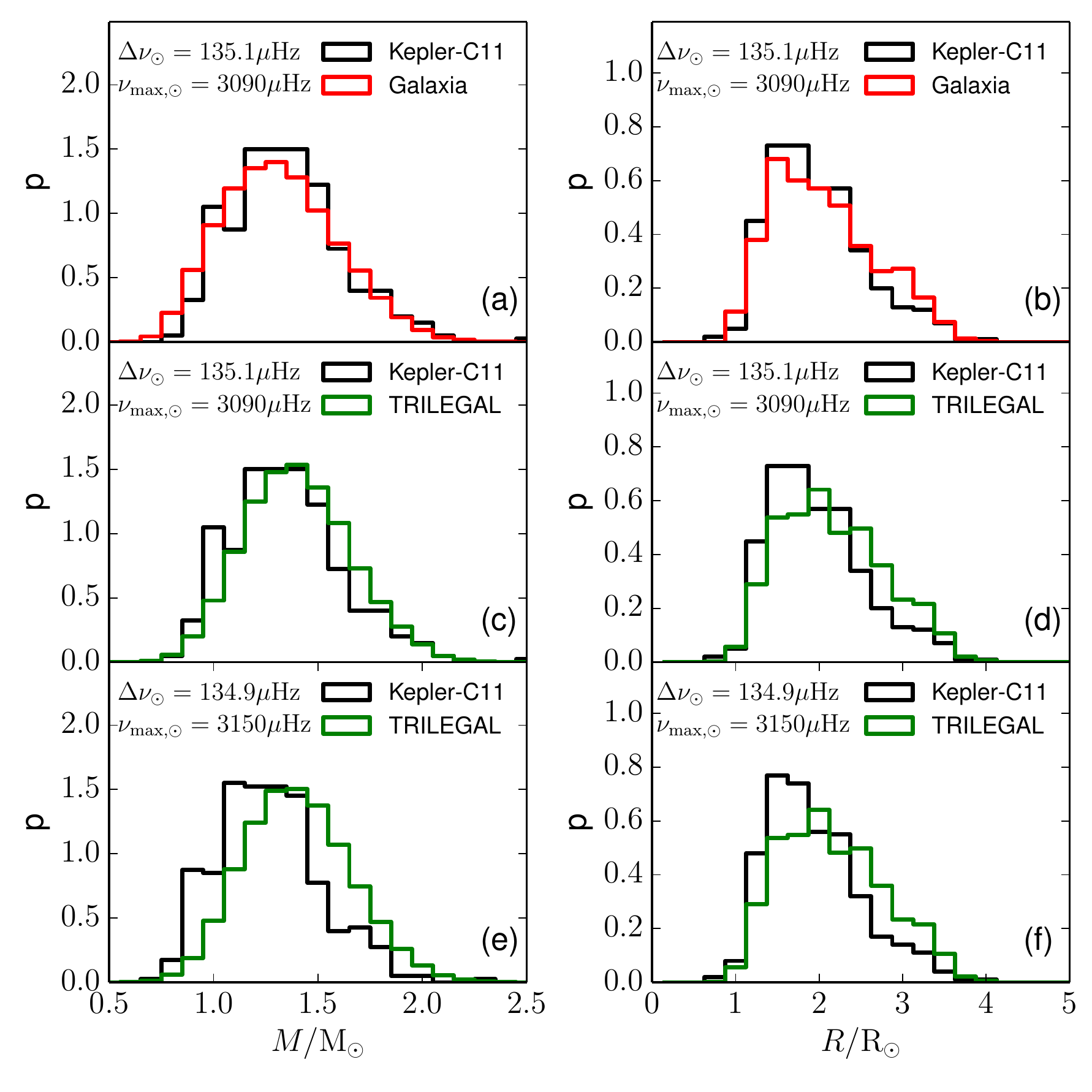}\caption{Same
  as \fig{c14s0_mr}, but for the Chaplin-11 sample.
\label{fig:c11s0_mr}}
\end{figure}
\section{Analysis of asteroseismic information} 

To compare the predictions of the Galactic model with the
asteroseismic information from {\it Kepler}, we
generated a new synthetic population of stars using both
{\sl Galaxia} and TRILEGAL. The synthetic stars were then
selected to match the selection criteria of the observed
stars. 
The main selection criteria was based on apparent
magnitude and a lower limit on $\nu_{\rm max}$. However, not all
targeted stars showed oscillations.
A scheme to compute the detection
probability was presented in \citet{2011ApJ...732...54C}
and this was used by \citet{2011Sci...332..213C} to select
stars from a Galactic model. In this scheme,
mass, radius, and effective temperature of each synthetic star were used to
predict its total mean oscillation power and the granulation noise in the
power spectrum.  The apparent magnitude was used to compute the instrumental
noise in the power spectrum which, combined with granulation noise, gave the
total noise.  The mean oscillation power and the total noise were then used
to derive the probability of detecting oscillations, $p_{\rm detect}$,
with less than 1\% possibility of false alarm.  Stars with $p_{\rm
detect}> 0.9$ were assumed to be detectable.
Hereafter we refer to this selection function as S0.

We applied the S0 selection function to the synthetic stars
and calculated the mass and radius distributions (\fig{c14s0_mr}).
\figs{c14s0_mr}a,b show the
predictions of {\sl Galaxia}, which match well with the
observations. \fig{c14s0_mr}c,d show
predictions of the TRILEGAL model,  which was used by
\citet{2011Sci...332..213C}.  Here, the predicted
mass distribution is shifted slightly to the right.
\fig{c14s0_mr}e,f also show the predictions of
TRILEGAL alongside observed stars, but with masses of
observed stars computed using the solar reference values adopted by
\citet{2011Sci...332..213C} in their analysis. As expected, this result is
the same as that presented by \citet{2011Sci...332..213C},
with a significant shift of the mass distribution
from TRILEGAL compared to the observed distribution.
In \fig{c11s0_mr}, we show the same analysis but using the
Chaplin-11 sample, whose size is slightly smaller than
the Chaplin-14 sample. As expected, 
we see the same trends as seen in \fig{c14s0_mr}.

The value of
$\nu_{\rm max, \odot}$ used by \citet{2011Sci...332..213C} is about 2\%
higher than the value used by us (\fig{c14s0_mr}a-d)
and this reduces the masses of the observed
stars by about 8\%, which exacerbates the mass discrepancy between the
TRILEGAL prediction and the observations.
The small difference in $\Delta \nu_{\odot}$, has no
significant effect.
To conclude, the mismatch in the stellar mass distribution
found by \citet{2011Sci...332..213C} can be alleviated if:
(i) one adopts a value of $\nu_{\rm max,\odot}$ that is
slightly smaller and (ii) we use {\sl Galaxia}
(with default settings) instead of TRILEGAL (with
default settings) as the Galactic model. 
This does not necessarily mean that the {\sl Galaxia} model is 
correct. The mass distribution is sensitive to the choice 
of the selection function and a bias in mass due to
an incorrect choice of the selection function can cancel 
out a bias due to an  
incorrect galactic model. Hence, in the next subsection,  
we investigate the accuracy of the selection function.

\begin{figure}
\centering \includegraphics[width=0.46\textwidth]{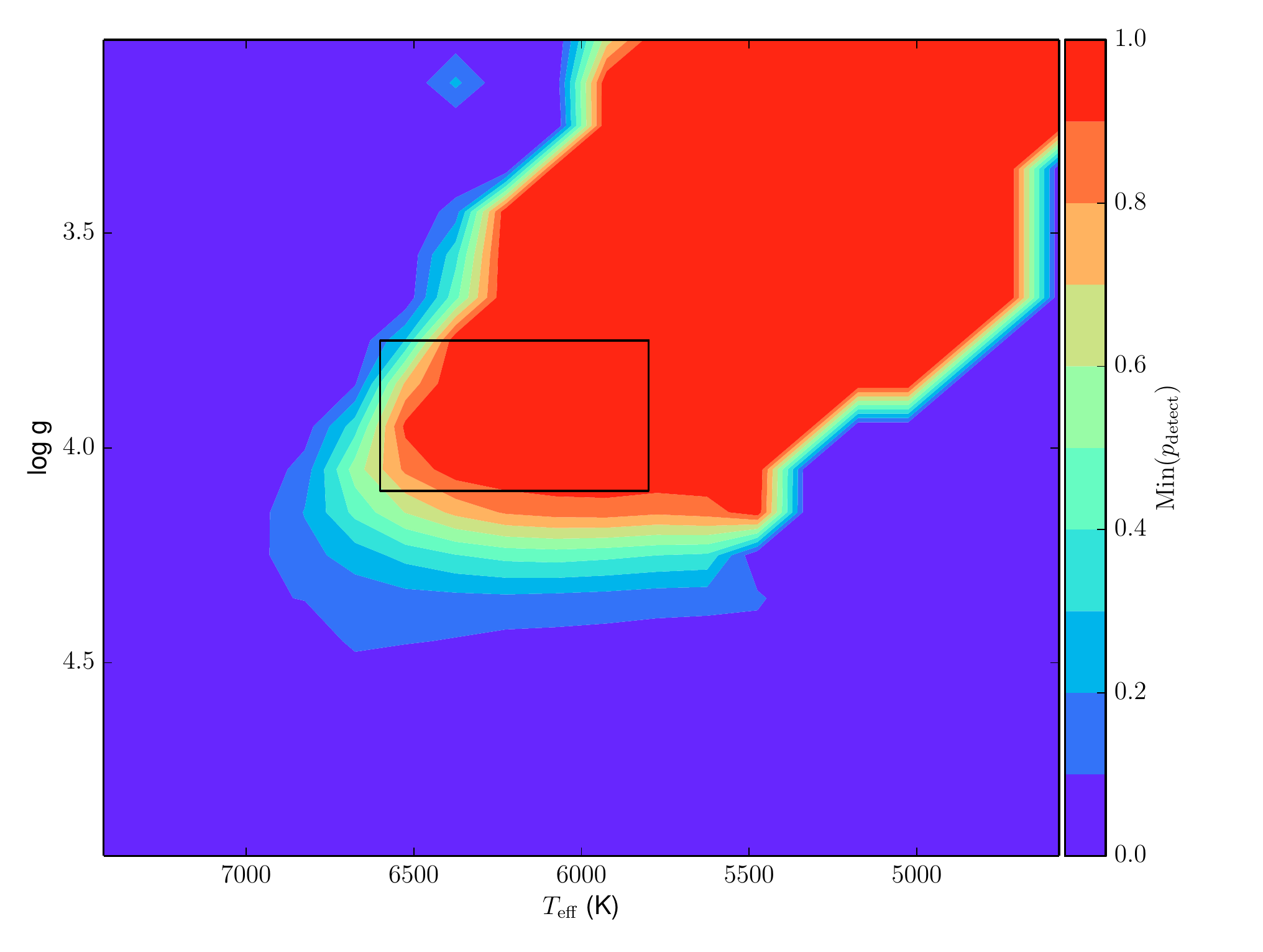}\caption{Minimum probability of detecting
oscillations as a function of $\log g$ and $T_{\rm
eff}$. The map is for  stars with apparent  magnitude
$r=11$ and is based on predictions from the code {\sl
Galaxia}. Detection probability is computed using
the scheme of \citet{2011ApJ...732...54C}.
A sharp transition from high to low detection
probability can be seen.
The rectangular  box marks the region where the sample is approximately
complete.
\label{fig:predict_2}}
\end{figure}

\begin{figure}
\centering \includegraphics[width=0.46\textwidth]{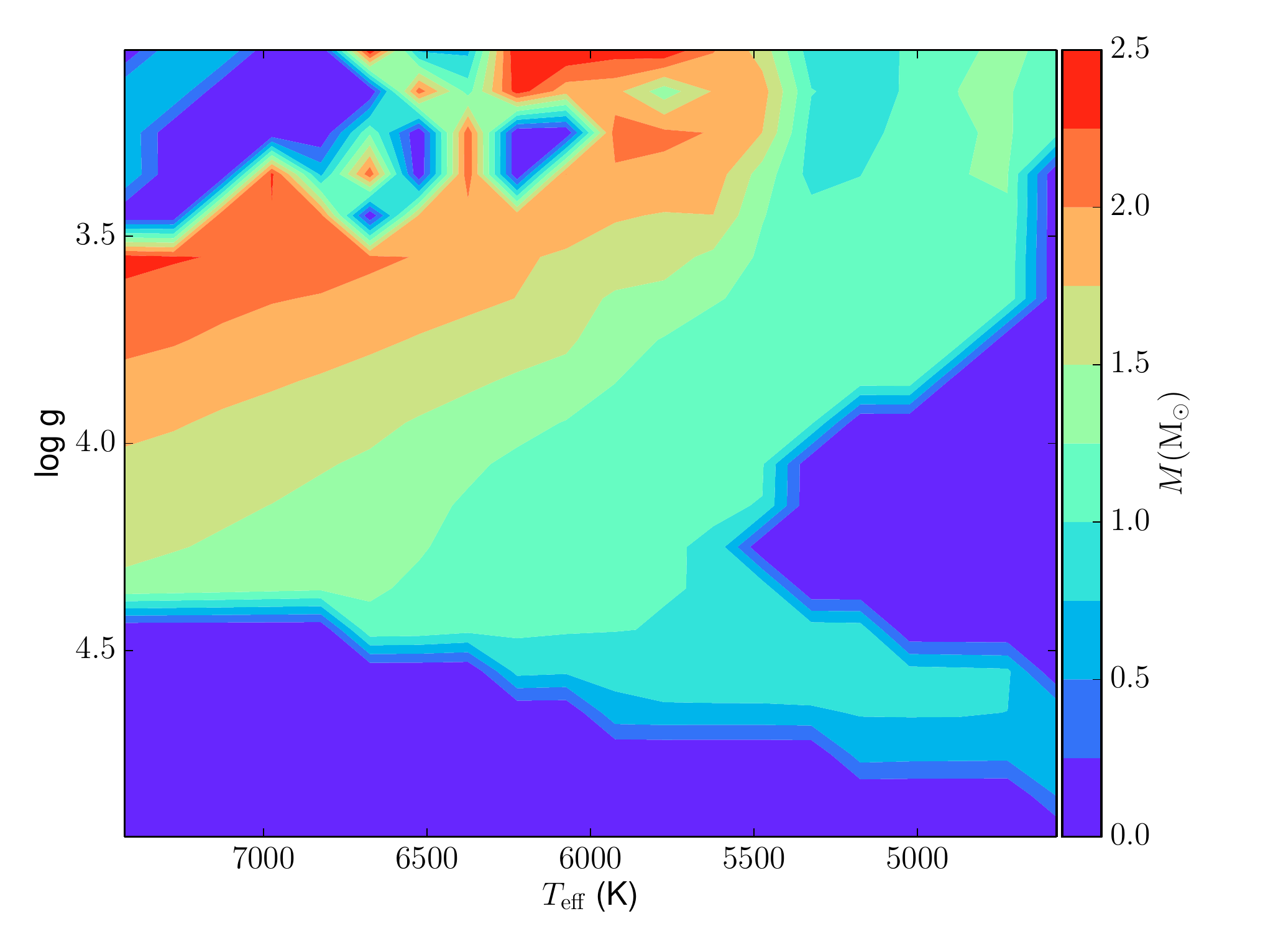}\caption{Mean mass of a star as a function of $\log g$ and $T_{\rm
eff}$, as predicted by {\sl Galaxia}, for  stars with
apparent  magnitude $r=11$ . The
mean mass increases as one moves diagonally from the lower right
to the upper left.
\label{fig:m_logg_teff}}
\end{figure}

\begin{figure}
\centering \includegraphics[width=0.46\textwidth]{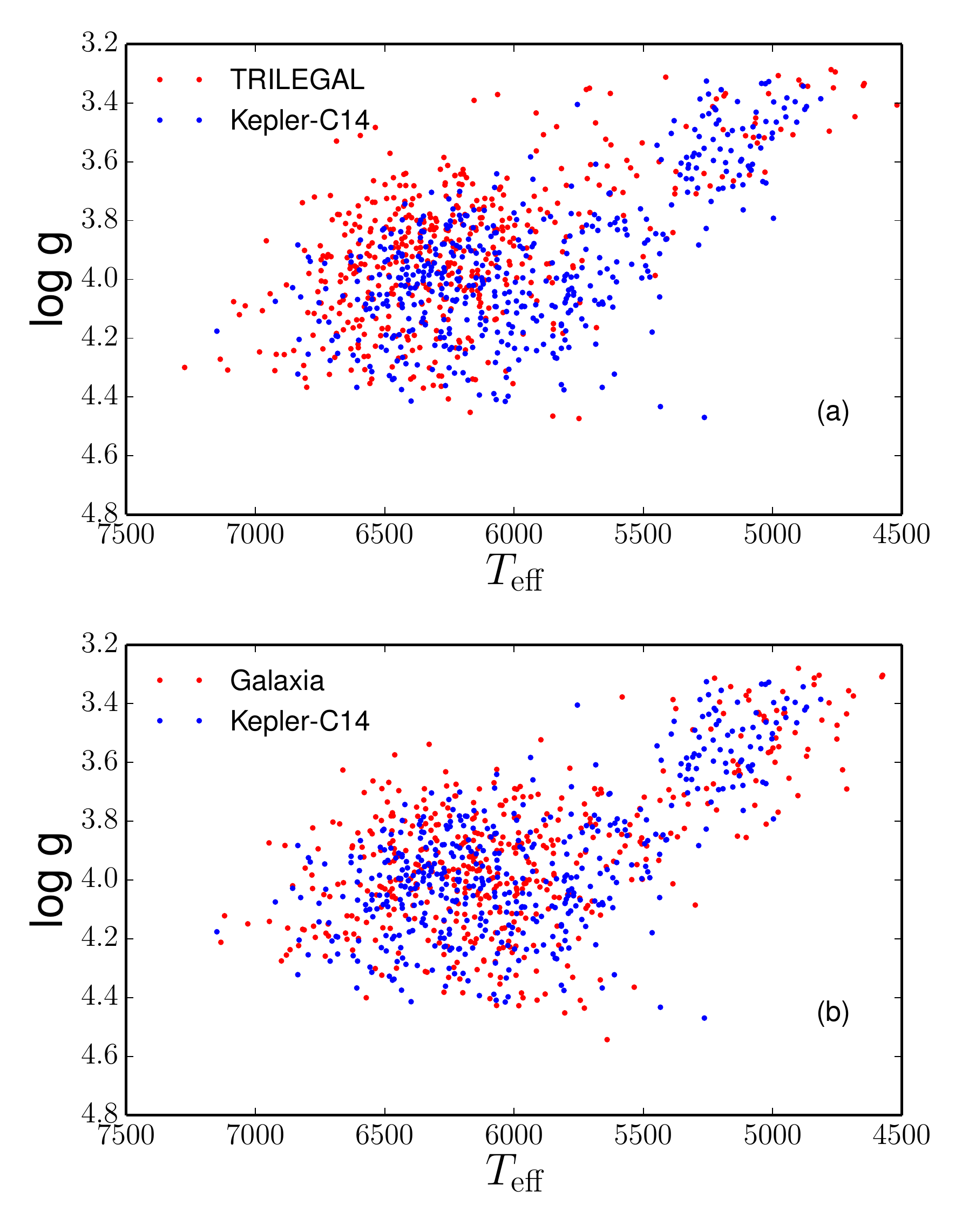}\caption{Stars  in $(\log g,T_{\rm
eff})$ space. The {\it Kepler} dwarfs and subgiants are shown as
blue points while synthetic stars with $p_{\rm detect}>0.9$ are
shown as red points. The observed data has an uncertainty of 
150 K in $Te_{\rm eff}$ and 0.2 dex in $\log g$, 
The simulated stars were convolved with 
uncertainties typical of that in the observed data,  
150 K in $T_{\rm eff}$ and 0.2 dex in $\log g$.  
Both Galactic models overpredict the
number of stars at around $(6500,3.7)$. In addition 
TRILEGAL underpredicts the number of stars at around $(5700,4.1)$ 
while {\sl Galaxia} predicts more stars at the 
low temperature end.
\label{fig:logg_teff_1}}
\end{figure}

\subsection{The role of the selection function}
The accuracy of a selection function that is
based on detectability of oscillations
hinges upon our ability to accurately predict the mean oscillation
power. The mean oscillation power is computed from
the maximum mode amplitude and for this an empirical relation
is used, which is prone to inaccuracies and
can potentially bias the selection function.

We now study whether altering the assumptions behind the
selection function have any effect on
the distribution of stars in the $(\log g,T_{\rm    eff})$
space.
In \fig{predict_2}, we show $p_{\rm detect}$ as a function of
$\log g$ and $T_{\rm    eff}$ for stars with $r=11$ mag,
selected from {\sl Galaxia}. Stars with high
$p_{\rm  detect}$ are confined to the region in $(\log g, T_{\rm
eff})$ space shown in red.
For the red region, the right boundary
is because there are no stars to the right of that
boundary. The left and lower boundaries
are because the oscillation amplitude decreases
with increasing $\log g$ and $T_{\rm  eff}$.
If we lower the assumed amplitudes in the selection
function,  the red region would  shift upwards and
to the right. Increasing the
assumed noise in the selection function or increasing $r$
would  have a similar effect.
On average, the stellar mass
increase as one moves diagonally from the lower right
to the upper left (\fig{m_logg_teff}). Hence, a bias in
$(\log g,T_{\rm    eff})$ space will also lead to a bias
in the masses of the stars.

In \fig{logg_teff_1}, we show the distribution of
the {\it Kepler} sample  in $(T_{\rm
eff},\log g)$ space
alongside predictions from Galactic models based on
the S0 selection function. The observed and predicted 
distributions match well in the central region but
differences can be seen along the boundary. 
Both models overpredict the number of stars at 
around $(6400,3.7)$.  Additionally, TRILEGAL 
predicts fewer stars around $(5600,4.1)$, while  
{\sl Galaxia} predicts more cooler stars ($T_{\rm} < 5000
{\rm\ K}$).
At the low-temperature end for evolved stars we
expect the detection probability to be high.
Hence, the differences seen here are most likely due to inaccuracies
in the models.
At high temperature, the detection
probability is low and is sensitive to the assumptions
made in the selection function.
Here, stars are close to the
instability strip where convection zones are thin, which
makes it difficult to model the mode driving and damping mechanisms.
Hence, the differences seen  here are most likely due to
inaccuracies in the relation used to predict mode
amplitudes.

The fact that the observed and the predicted distributions
of stars do not match in $(T_{\rm  eff},\log g)$ space
means the mass distributions will also not match.
To eliminate this bias, we created new selection functions
by resampling the model stars to
satisfy the observed distribution of stars in
$(T_{\rm  eff},\log g,r)$ space.
The disadvantage of such resampling
is that we reduce our sensitivity to model-based
differences, because the first order differences are already
taken out. However, they are still useful to understand
systematics related to asteroseismic analysis.
Below are two ways for devising such new  selection
functions.
\begin{itemize}
\item S1: Here we resample the model stars to match the distribution of
observed stars in $(T_{\rm eff}, \log g,r)$ space. 
This is done by dividing the $(T_{\rm eff}, \log g,r)$ space
into bins and then making sure that each bin 
has the same number of model and observed stars. Because the 
number of observed stars is low,    
one has to adopt large bins and this can affect  
the accuracy of the selection function. 
\item S2: In this we select a box in $(T_{\rm  eff},\log g,
r)$ space where we expect the
detection probability to be close to 1 and where the distribution
of the observed data
matches that of the Galactic model. Compared to the case S1, 
the case S2 leads to fewer stars, 
but has a more accurate selection function. 
The selected box,
\be
p(S|T_{\rm eff}, \log g, r)=
\begin{cases}
1 & \text{if } (5800<T_{\rm
eff}<6600)\& \\
& (3.8<\log g <4.1)\& \\
& (r<11) \\
0 & \text{otherwise},
\end{cases}
\ee
is shown in \fig{predict_2}, and it can be seen that
the box is mainly inside the region of high detection
probability (red).
\end{itemize}

\begin{table}
\caption{Value of $f_{\nu_{\rm max}}$ for which the
distribution of observed
masses matches best with the masses predicted by the
Galactic models. The values in square bracket are for the
case when the $f_{\Delta \nu}$ correction factor is
applied to the synthetic stars. }
\begin{tabular}{l l l} \hline
Selection& $f_{\nu_{\rm max}}$ {\sl Galaxia} & $f_{\nu_{\rm
max}}$ TRILEGAL\\
Function & & \\ \hline
S0 & $1.006[1.004]\pm 0.002$ & $0.982[0.981]\pm 0.002$\\
S1 & $1.018[1.015]\pm 0.002$ & $1.002[1.001]\pm 0.002$\\
S2 & $1.019[1.011]\pm 0.004$ & $1.004[0.994]\pm 0.004$\\ \hline
\end{tabular}
\label{tab:tb2}
\end{table}

\begin{figure}
\centering \includegraphics[width=0.47\textwidth]{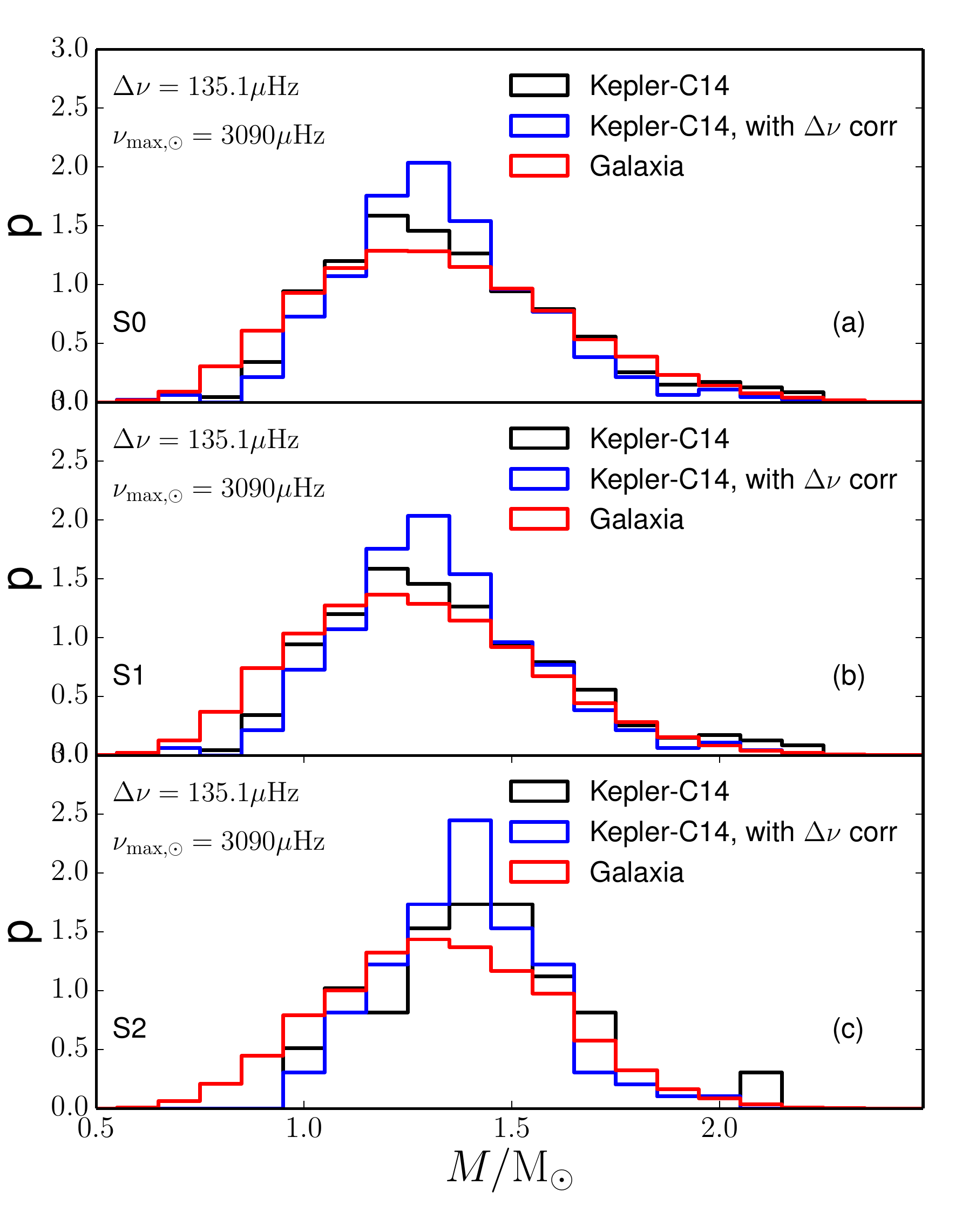}\caption{Mass distribution of observed stars
alongside predictions from {\sl Galaxia}, using three different
selection functions. We use the same solar reference values
for all three cases. The blue line is for the case where
theoretically predicted corrections to $\Delta \nu$ scaling
relation are applied. The corrections were computed assuming
solar metallicity ($Z=0.019$) for all the observed stars.
\label{fig:galaxia_trilegal_fm_sigmam2}}
\end{figure}

In \fig{galaxia_trilegal_fm_sigmam2}, we compare the
predicted mass distributions with observations, for
the three different selection functions, S0, S1 and S2.
Compared to the observed distributions, the predicted
distributions are shifted toward lower masses and are
slightly broader. For S0 the shift is minimal, but for
S1 and S2 it is significant.
We note that \fig{galaxia_trilegal_fm_sigmam2}a (S0)
is the same as \fig{c14s0_mr}a.
To quantify the shift between the mass distributions,
we determined the value of $f_{\nu_{\rm max}}$ that
minimizes the Kolmogorov-Smirnov statistic between the
observed and predicted distributions. The uncertainty on the
estimate was computed using bootstrapping.
The results are shown in \tab{tb2}.
The two selection
functions (S1 and S2) that are based on observables
like gravity and temperature, give similar values for
$f_{\nu_{\rm max}}$, but the value is different for S0.
It is clear that the selection function plays a crucial role and
can bias $f_{\nu_{\rm max}}$ by 2\%. The difference
between TRILEGAL and {\sl Galaxia} is also about 2\%.

Detailed theoretical modelling of oscillations shows 
that there are departures from the $\Delta \nu$ scaling
relation that depend upon metallicity, mass, and age.
As mentioned in \sec{astero}, we accommodated
these departures using the correction factor $f_{ \Delta \nu}$.
We computed these corrections using the
publicly available code ASFGrid\footnote{\url{http://www.physics.usyd.edu.au/k2gap/Asfgrid}}.
ASFGrid  uses the code MESA (v6950)
\citep{2011ApJS..192....3P,2013ApJS..208....4P,2015ApJS..220...15P}
for stellar evolution and the code GYRE
\citep{2013MNRAS.435.3406T} for deriving oscillation frequencies.
The correction factors are computed as a function of
metallicity, mass, and age,  see \citet{2016ApJ...822...15S}
for further details.
Applying these
corrections makes the observed mass distribution
narrower (blue lines in \fig{galaxia_trilegal_fm_sigmam2}).
This is expected because for the stars that we study,
the mean mass of a star increases with temperature
and the correction factor decreases with temperature.
The combined effect
is that for high-mass stars, 
the mass decreases, and for low-mass stars, the mass
increases. This leads to narrowing of the overall
distribution. Computing corrections requires
metallicity, and we adopted the spectroscopic  metallicities
reported  by \citet{2015ApJ...808..187B}.
Instead of applying the correction to the observed stars, 
we can also apply the reciprocal correction to
the synthetic stars whose metallicity is known exactly.
Doing so had negligible effect on the values
quoted in \tab{tb2} (see results in square brackets).

\section{Discussion and conclusions}
We have compared the asteroseismic properties of dwarfs and
subgiants observed by Kepler against predictions of two population
synthesis models of the Galaxy, TRILEGAL and {\sl Galaxia}.
The previous study by \citet{2011Sci...332..213C}
using TRILEGAL found that
stellar population synthesis based models
overestimated the number of high-mass stars, which
we are able to reproduce.
We identified three potential factors that can shift the
model mass distributions toward higher masses relative to
the observed masses.
First, TRILEGAL most likely overpredicts the number
of young massive stars as it fails to match the $J-Ks$ color
distribution of stars in 2MASS \citep{2016ApJ...822...15S}.
Second, we found that a choice of $\nu_{\rm max,\odot}$ that is 2\%
lower than that adopted by \citet{2011Sci...332..213C},
which  is equally valid given the uncertainty in
the actual value, can increase the
observed masses by about 6\%.
Finally, we found that if a selection function based on oscillation
amplitudes is used, the Galactic models cannot reproduce
the distribution of the observed sample in $(\log g,T_{\rm
eff})$ space.
This might be due to inaccuracies in the model, but
could also be due to inaccuracies in the assumed selection
function.
Selecting the synthetic stars to satisfy the
distribution in $(\log g,T_{\rm eff})$ space removed this
bias but shifted the model mass distribution to lower masses
(\fig{galaxia_trilegal_fm_sigmam2}b,c).

The bias due to the mismatch of
the color distribution can be corrected by using a model
such as {\sl Galaxia}, which does not show such a mismatch.
The bias due to inaccuracies in
the selection function based on oscillation amplitudes can be
reduced by using a selection function
based on $\log g$ and $T_{\rm eff}$  of the observed
sample. Doing so, we find that the
mass distribution of {\sl Galaxia}
is shifted toward lower masses
and is also slightly broader compared to the observed
distribution.
A similar effect was also seen for
the {\it Kepler} red giant sample \citet{2016ApJ...822...15S},
so the underlying cause
might be the same.
Applying corrections to the $\Delta \nu$
scaling relation predicted by stellar models makes the observed
mass distribution narrower than observed  but does not change
the mean.
The disagreement in the mass distributions  reported here,
translates to about 2\% change in $\nu_{\rm
max}$, which is comparable  to
the current uncertainty in the $\nu_{\rm max}$ scaling relation.
In future, we need to verify the scaling relations to better
than 2\% to put better constraints on the Galactic models.

However, {\sl Galaxia}  failed to match the distribution
of observed stars in $\log g$ and $T_{\rm eff}$ space.
This also needs to be investigated in future.
The mismatch at high $T_{\rm eff}$
could be due to inaccuracies in predicting
oscillation amplitudes because the detection probability
of a star in this region is sensitive to its
assumed amplitude.
However, the mismatch at low $T_{\rm  eff}$ is most
likely due to inaccuracies in the model,
because here we expect the detection probability to be close to
1. Parallaxes, and hence luminosities, from Gaia will
help resolve this issue because luminosity correlates
with gravity.

\section*{Acknowledgements}
We acknowledge the support of {\it Galactic
Archaeology and Precision Stellar Astrophysics} program
organized by  Kavli Institute for Theoretical Physics
(National Science Foundation Grant No. NSF PHY11-25915)
for facilitating helpful discussions of results in this
paper.
We thank William
Chaplin for allowing us to use his code for computing
the probability of detecting oscillations.

S.S. is funded through ARC DP grant 120104562
(PI Bland-Hawthorn) which supports the HERMES project.
D.S. is funded through Future Fellowship from the Australian
Research Council (ARC).
J.B.H. is funded through Laureate Fellowship from the Australian
Research Council (ARC).
D.H. acknowledges support by the Australian Research
Council's Discovery Projects funding scheme
(project number DE140101364) and support by the
National Aeronautics and Space Administration under
Grant NNX14AB92G issued through the Kepler Participating
Scientist Program.

\bibliographystyle{apj}
\bibliography{}

\end{document}